\documentclass[10pt,conference,letterpaper]{IEEEtran}

\usepackage{amsfonts}
\usepackage{amsmath}
\usepackage{amssymb}
\usepackage{color}
\usepackage{latexsym}
\usepackage{caption}
\usepackage{subcaption}
\usepackage{graphicx}
\usepackage{setspace}
\usepackage{multirow}
\usepackage{cite}
\usepackage{enumerate}
\usepackage{enumitem}
\usepackage{tabularx}
\usepackage{url}
\graphicspath{{}}

\newcommand{\cutoff}[1]{}
\begin{document}

\title{Real-time Video Streaming on mmWave Communication Testbed}

\title{A Real-Time mmWave Communication Testbed with Phase Noise Cancellation\vspace{-0.15in}}

\author{\IEEEauthorblockN{Adnan Quadri\IEEEauthorrefmark{1},
		Huacheng Zeng\IEEEauthorrefmark{1}
		and
		Y. Thomas Hou\IEEEauthorrefmark{2}
		}	
	\IEEEauthorblockA{
		\IEEEauthorrefmark{1}Department of Electrical and Computer Engineering, University of Louisville\\
		\IEEEauthorrefmark{2}Department of Electrical and Computer Engineering, Virginia Polytechnic Institute and State University\\
		Email: adnan.quadri@louisville.edu,
		huacheng.zeng@louisville.edu, 
		thou@vt.edu}\vspace{-0.25in}
	}

\maketitle

\begin{abstract}
As the spectrum under 6 GHz is being depleted, pushing wireless communications onto millimeter wave (mmWave) frequencies is a trend that promises multi-Gbps data rate. 
mmWave is therefore considered as a key technology for 5G wireless systems and has attracted tremendous research efforts. 
The booming research on mmWave necessitates a reconfigurable mmWave testbed that can be used to prototype and validate new research ideas in real wireless environments. 
In this paper, we develop an easy-to-use mmWave testbed using commercial off-the-shelf devices (USRP and 60 GHz Tx/Rx RF frontends) and open-source software package (GNU Radio).
A key component of our testbed is a phase noise cancellation (PNC) scheme, which can significantly reduce the phase noise at the receiver by leveraging the pilot signal inserted at the transmitter. 
We have implemented a simplified version of IEEE 802.11 PHY on this mmWave testbed.
Experimental results show that, with the PNC scheme, our testbed can achieve -20 dB EVM data transmission for real-time video streaming.


\end{abstract}

\section{Introduction}

As millions of mobile devices are introduced into the market every year, wireless data traffic has been growing at a rate of 50\% per year and this trend is expected to accelerate in the next decade with the proliferation of video-centric applications and the promotion of Internet-of-Things (IoT) systems \cite{gubbi2013internet}. 
To address the unprecedented traffic demand, novel wireless technologies are being investigated towards revolutionizing the existing wireless infrastructure, with the objective of improving the wireless network capacity by orders of magnitude. 
Given that the spectrum under 6 GHz is fundamentally limited (and already very crowded), wireless communications on millimeter wave (mmWave) frequencies are widely regarded as a promising solution as they can offer very large license-free bandwidth (spanning 57 GHz to 64 GHz in many countries) for communication purpose and enable high-dimensional MIMO operations, thanks to the small wave-length of mmWave frequencies.  
\cutoff{
Recognizing its huge potential, communications research community and telecommunications companies widely regard mmWave as an enabler for next-generation wireless access networks \cite{rappaport2013millimeter}. 
This is evidenced by recent advancement of energy-efficient 60 GHz transceivers from semiconductor companies (e.g., Qualcomm and Intel) as well as the pushing force toward the commercialization of low-cost mmWave sensing devices for medical, security, imaging, localization, and tracking applications (see, e.g., \cite{wei2015mtrack,zhu2015reusing}). 
}

For wireless networking research, experimentation plays a very important role. 
Although theoretical analysis and computer-aided simulation are important tools in studying signal modulation techniques and analyzing network protocols, it is essential that new research ideas be implemented and assessed on wireless testbeds to examine their performance in real-world environments. 
The prosperity of mmWave networking research necessitates an easy-to-use reconfigurable mmWave testbed that provides researchers with access to the low layers (PHY and MAC).
Ideally, such an experimental testbed should allow researchers to customize coding and modulation schemes at the PHY layer, craft protocols and scheduling algorithms at the MAC layer, and prototype new mmWave-based applications in a convenient manner.
The necessity and importance of such a mmWave testbed are reflected by the success of its microwave counterparts such as USRP \cite{ettus} and WARP \cite{khattab2008warp}\cutoff{, and Sora \cite{tan2011sora}}, which have reshaped the landscape of wireless experimentation in the past decade and led to the development of many important wireless technologies.

Realizing the importance of mmWave testbeds, pioneering efforts have been invested to develop such testbeds for research use \cite{saha2019x60}\cite{zhang2016openmili}\cite{zetterberg2015open}\cite{arnold2015spectrum}. 
In \cite{saha2019x60}, a programmable testbed (called X60) was developed to support mmWave PHY and MAC design. 
However, such a testbed is highly costly ($\ge$ \$200,000) and not affordable for most research groups.
In \cite{zhang2016openmili}, a reconfigurable platform called OpenMili was developed for broadband mmWave communications.
However, it involves FPGA programming, which may pose a technical barrier for some researchers.
In \cite{zetterberg2015open}, a mmWave testbed was built using USRP N210 and 60 GHz Tx/Rx RF frontends.
This mmWave testbed addresses the phase noise problem through hardware reclocking, which reduces the flexibility of the testbed.

In this paper, we develop an easy-to-use software-defined mmWave testbed using commercial off-the-shelf (COTS) devices and open-source software package, in the hope that such a testbed can be easily reproduced by other research groups to facilitate their experimental mmWave research. 
From the hardware perspective, our mmWave testbed integrates USRP devices from National Instruments, RF Single-Ended to Differential Converter, 60 GHz RF modules and 60 GHz 42 dBi horn antennas from Pasternack. 
From the software perspective, our testbed takes advantage of widely-used GNU Radio software packet in Linux for the implementation of PHY-layer signal processing and MAC-layer protocols. 
It eases the process of prototyping by using C++ or Python for algorithm implementation and using Gnuradio-Companion for GUI control.
Such a configuration eliminates the need for knowledge of FPGA programming and facilitates the experimentation process.


A key component of our testbed is phase noise cancellation (PNC). 
Phase noise is a notorious problem for mmWave systems.
It is caused during the generation of the extremely high oscillation carrier frequency. 
Different from existing testbeds, which address the phase noise problem through hardware reclocking (using external high-precision clock source) \cite{zhang2016openmili,zetterberg2015open}, our testbed resorts to a software solution by developing a practical scheme to cancel the phase noise in real time. 
As a result, there is no need for hardware modification on the COTS devices. 
The removal of hardware reclocking not only simplifies hardware complexity and reduces hardware cost, but also improves the operational flexibility of experimentation on our testbed. 

We summarize the features of our mmWave testbed as follows:
(i) Our mmWave testbed was built using COTS devices, which is easily reproduced by other research groups to facilitate their experimental mmWave research.
(ii) Our mmWave testbed does not require hardware reclocking. Instead, it comes with a lightweight phase noise cancellation scheme to mitigate the phase noise for signal detection on the receiver side. 
(iii) GNU Radio software package is used for the implementation of signal modulation, protocols, and algorithms. 
Researchers with experience in USRP and GNU Radio will find this testbed easy to use. 
(iv) Our mmWave testbed can support real-time video streaming with about \mbox{-20} dB EVM. 
A demo of video streaming on our mmWave testbed can be found in \cite{videodemo}.


\section{A Real-Time mmWave Testbed}

%

\begin{table}
	\centering
	\caption{The list of hardware components used for the implementation of our mmWave testbed.}
	{\fontsize{6}{8}\selectfont	
	\begin{tabular}{|l|l|l|}
		\hline
		\!\!\!\!\!\! Item Name& \!\!\!Item Description    &  Price \\ \hline
		\!\!\!\!\!\! PC (2) & \!\!\!Baseband signal processing \& video streaming    &  \!\!\!\!\$1,000\!\!\!\!\!\!\\ \hline
		\!\!\!\!\!\! USRP N210 (2)& \!\!\!Convert digital baseband signal to analog IF signal \!\!\!\!\!\! & \!\!\!\!\$3,886\!\!\!\!\!\!\\ \hline
		\!\!\!\!\!\! RF Single-Ended to Diff. Converter (4) \!\!\!\!\!\!& \!\!\!Interface match for I and Q signals    & \!\!\!\!\! \$56 \!\!\! \\ \hline
		\!\!\!\!\!\! 30 dB Attenuator (2)    & \!\!\!To match 60 GHz system's voltage input requirement\!\!\!\!\!\!&  \!\!\!\!\$60 \!\!\!\!\!\! \\ \hline
		\!\!\!\!\!\! 60 GHz Tx/Rx Circuit Boards (1) & \!\!\!Up-/down-conversion for 60 GHz RF signals   & \!\!\!\!\$12,383 \!\!\! \!\!\! \!\!\!\!\!\! \\ \hline
		\!\!\!\!\!\! Horn Antenna (2)   & \!\!\!Highly directional antenna with 42 dBi gain&\!\!\!\!\$4,354 \!\!\!\!\!\!\\ \hline
	\end{tabular}
	}\vspace{-0.2in}
\label{tab:hardware}
\end{table}

In this section, we first describe the hardware devices and the software configuration of our mmWave testbed, and then present some experimental results obtained from the testbed. 

\subsection{Hardware Setup}
\begin{figure}
	\centering
	\includegraphics[width=3.5in]{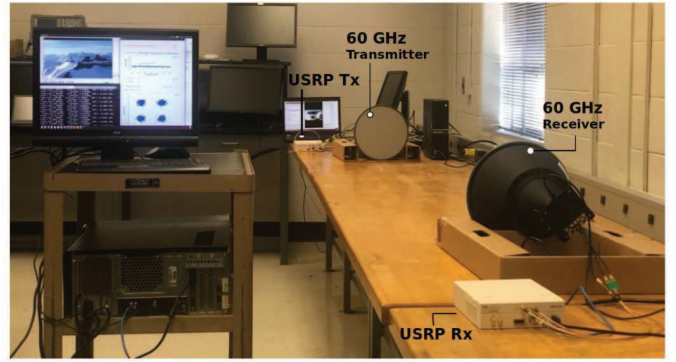}
	\caption{Testbed setup for mmWave communication.}\vspace{-0.2in}
	\label{fig:practiaclsettingmmwave}
\end{figure}

\begin{figure}
	\centering
	\includegraphics[width=3.5in]{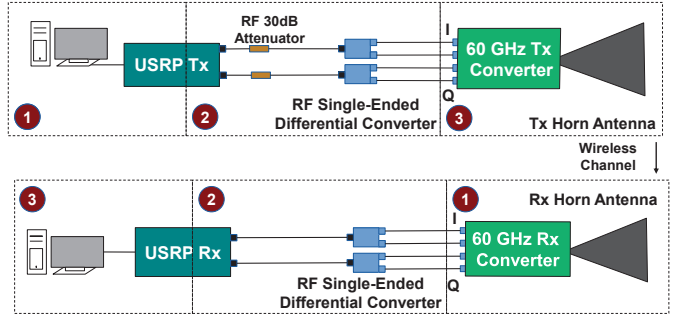}
	\caption{Schematic diagram of our mmWave testbed.}\vspace{-0.2in}
	\label{fig:mmwavetestbedsetup}
\end{figure}

Fig.~\ref{fig:practiaclsettingmmwave} shows a photo of our hardware setup and Fig.~\ref{fig:mmwavetestbedsetup} shows the schematic diagram of hardware components.
As shown in the figures, the implementation of the mmWave testbed uses the following components:
Tx/Rx computers, USRP devices with BasicTx/RX daughterboards \cite{ettus}, 30 dB RF attenuators\cite{ettus}, single-ended to differential converters, Pasternack 60 GHz RF frontends \cite{mmwaveSys}, and proper types of cables for connection.  
The functionality and price of each hardware component is provided in Table~\ref{tab:hardware}.

\noindent \textbf{Transmitter-Side Setup.}
The host PC serves as the source of data bits.
It is connected to a USRP device (with BasicTx daughterboard) via a crossover Ethernet cable. 
The USRP device is configured to Tx mode and converts the digital signal to analog intermediate frequency (IF) signal, providing I (in-phase) and Q (quadrature) outputs. 
The single-ended I/Q outputs of USRP are then connected to the differential I/Q inputs of Pasternack 60 GHz Tx RF frontend, with two 30 dB RF attenuators and two single-ended to differential converters placed in the middle to match the interface and voltage levels.
The Pasternack 60 GHz Tx RF frontend circuit comes with a built-in oscillator.
Hence, there is no need for external oscillation clocking.
The 60 GHz RF frontend circuit converts the IF signal to mmWave signal, and the carrier frequency can be set by users.
The 60 GHz RF frontend circuit is connected with a horn antenna via WR-15 waveguide interface. 
The horn antenna has 42 dBi gain, making it possible to communicate over 500 meters.

\begin{figure*}
	\begin{subfigure}[b]{1.75in}
		\centering
		\includegraphics[width=1.75in,height=1.3in]{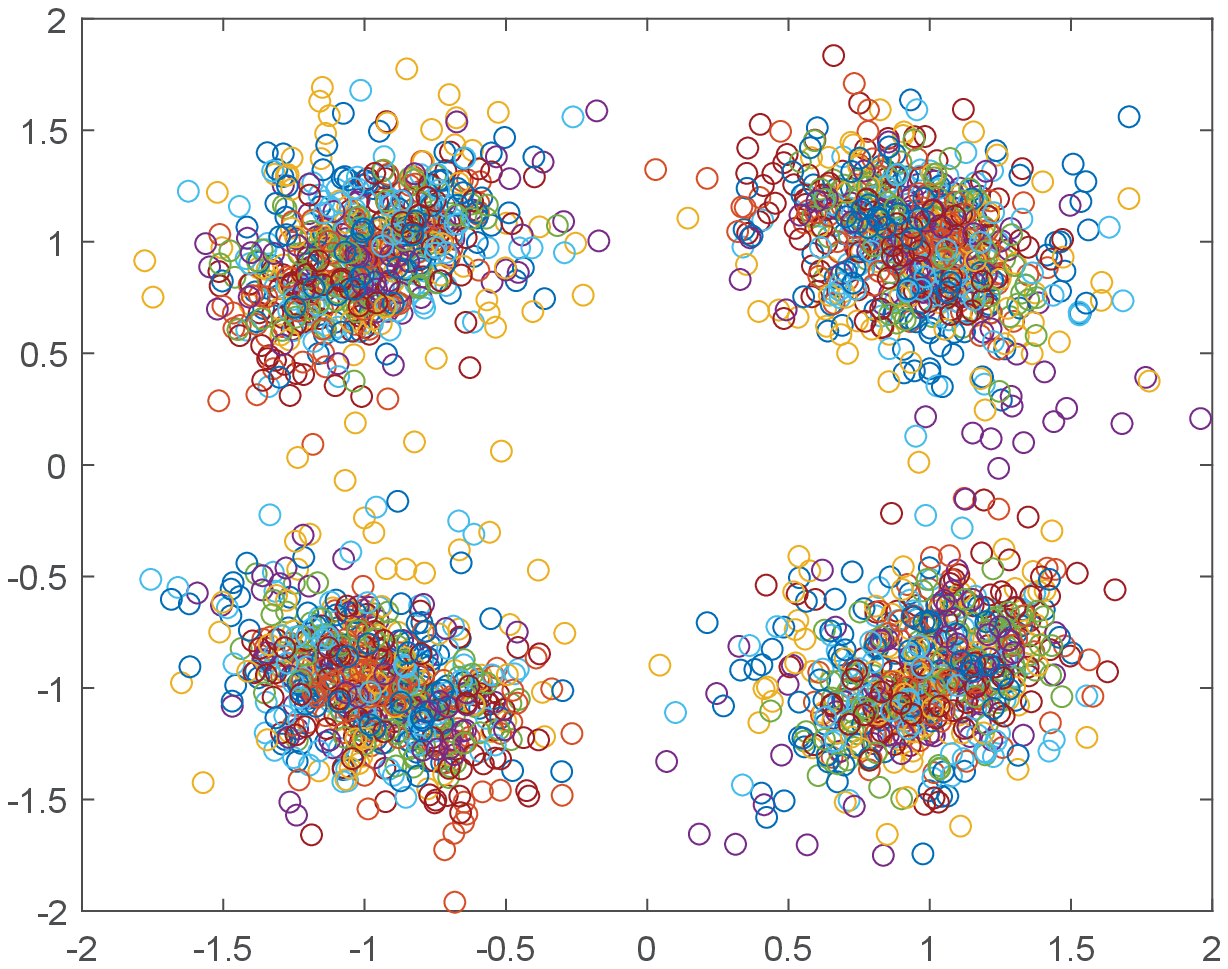}
		\caption{Constellation of decoded signal at the receiver (EVM = -8 dB).}
		\label{fig:constellation_wo_pnc}
	\end{subfigure}	
	~~
	\begin{subfigure}[b]{2.45in}
		\centering
		\includegraphics[width=2.35in,height=1.3in]{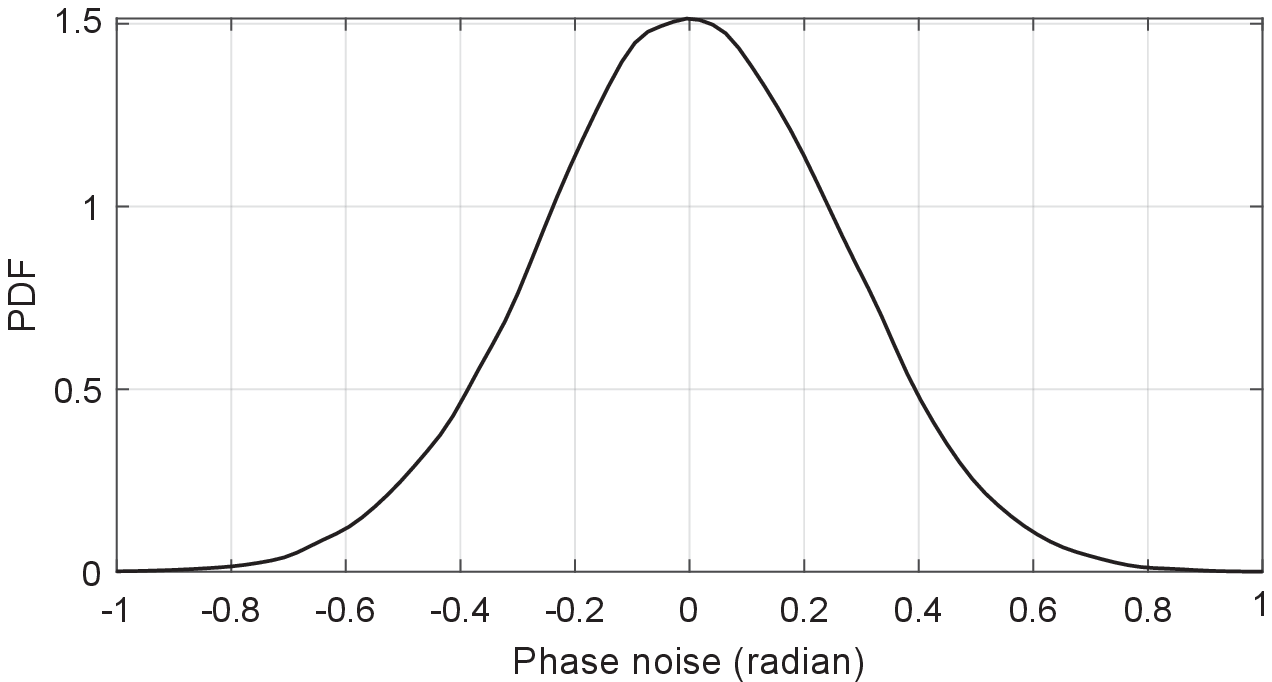}
		\caption{Probability density function (PDF) of phase noise.}
		\label{fig:phasenoisedis1}
	\end{subfigure}
	~~
	\begin{subfigure}[b]{2.15in}
		\centering
		\includegraphics[width=2.5in,height=1.3in]{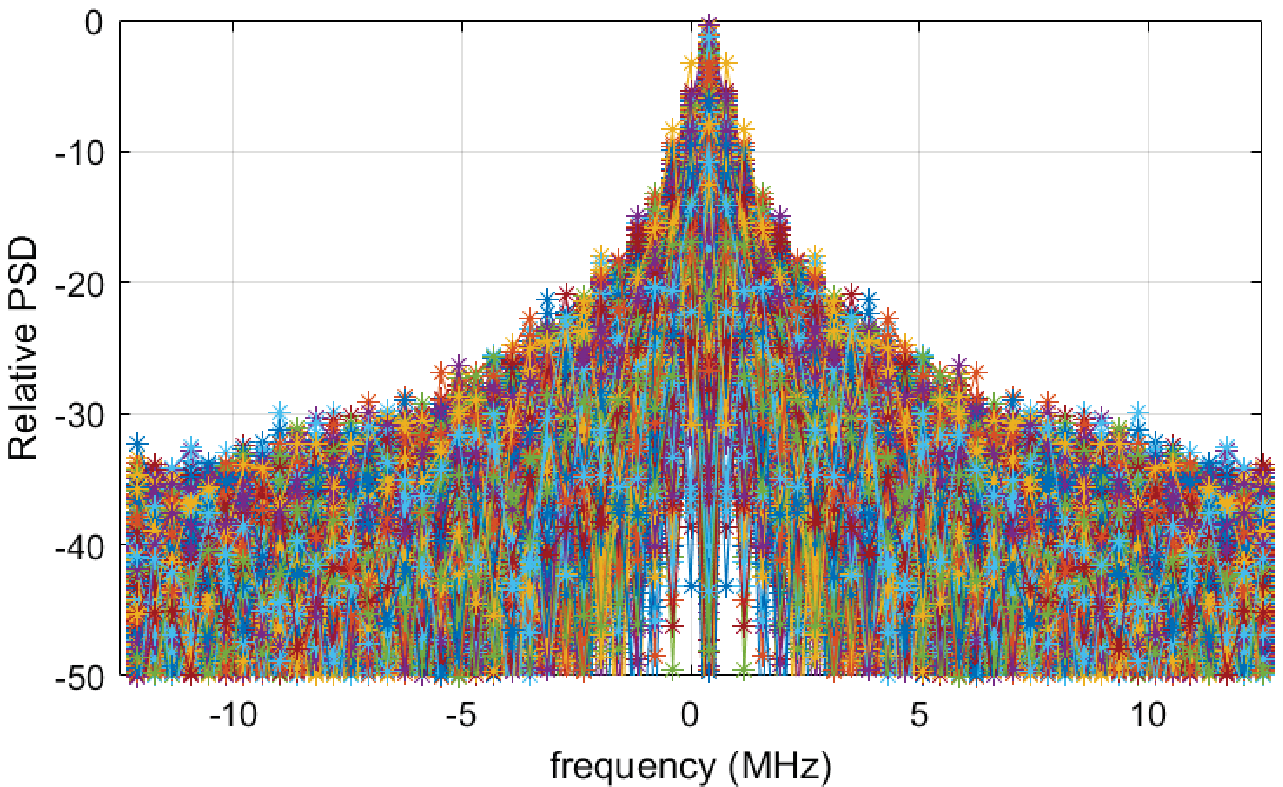}
		\caption{Power spectral density (PSD) of phase noise.}
		\label{fig:phasenoisedis2}
	\end{subfigure}
	\caption{\vspace{-0.05in}Experimental results measured on our mmWave testbed (without PNC).}\vspace{-0.15in}
	\label{fig:phasenoisedis}
\end{figure*}

\noindent \textbf{Receiver-Side Setup.}
The setup on the receiver side is an inverse connection of the transmitter side. 
As shown in Fig.~\ref{fig:mmwavetestbedsetup}, a horn antenna is connected to Pasternack 60 GHz RF frontend, which converts the RF signal to IF signal. 
The I/Q IF signals are then streamed to USRP through two differential to single-ended converters.
Note that, since the 60 GHz RF frontend circuit has baseband attenuation control, there is no need for 30 dB attenuators on the receiver side. 
The IF signal is then converted to the baseband signal by USRP and fed to the host PC via crossover Ethernet cable. 
The host PC performs baseband signal processing to recover the signal and decode the original data packet. 

\noindent \textbf{Frequency Range.}
The Pasternack 60 GHz Tx/Rx frontend circuit boards can up-/down-convert the signal to and from 60 GHz band. 
It can operate in frequencies ranging from 57.24 GHz to 64.80 GHz with a step of 0.54 GHz. 
The 60 GHz frontend circuit boards are connected with the horn antennas via WR-15 waveguide, which is a standard interface.
The horn antenna can be easily replaced with a phased array antenna if available.

\subsection{Software Setup}

Two software packages are used to control the testbed. 

\noindent
\textbf{Software for 60~GHz Frontend Control.}
Pasternack provides a software program to set up the carrier frequency and the attenuation values for the 60~GHz Tx/Rx RF Frontends via USB interface. 
The software provided by Pasternack can only run in Windows OS. 
But, through reverse engineering, the 60~GHz Frontend devices can be configured in Linux OS \cite{zhang2016openmili}.
By doing so, the processing of baseband signal and the control of RF frontends can be integrated into a unified system. 
For experiments, this software only needs to run one time to set up the parameters.
Those parameters will be stored in registers of the devices and will be automatically loaded when the devices are powered up next time.

\noindent
\textbf{GNU Radio Software Package.}
GNU Radio is a free software development toolkit that provides signal processing blocks to implement software-defined radios and signal-processing systems.
We installed GNU Radio on the two host PCs in Linux OS to perform baseband signal processing.
On the transmitter side, the host PC generates baseband signal, which is routed to USRP for frequency up-conversion via crossover Ethernet cable. 
On the receiver side, the host PC receives baseband signals from USRP and performs digital signal processing to decode the data packet.

In our testbed, GNU Radio Out Of Tree (OOT) Modules are used to build the signal processing blocks for mmWave communications. 
Besides the existing signal processing blocks, we have developed custom-designed OOT modules using C++ programming language for the mmWave testbed.
Particularly, the GNU Radio OOT modules require scientific computing to perform linear algebra operations such as FFT/IFFT, eigen decomposition, singular value decomposition (SVD), and matrix inversion.
To address this issue, we resort to Armadillo -- C++ library for linear algebra and scientific computing.
Our experimentation shows that Armadillo has a superior performance when running in real time for signal processing on both transmitter and receiver sides.

\subsection{Experimental Observations}
\label{sec:exp_obs}

\noindent \textbf{OFDM Transmission.}
On the mmWave testbed described above, we implemented a simplified version of IEEE 802.11n PHY to support video streaming from the transmitter to the receiver on 60~GHz frequency. 
We implemented modulation schemes such as BPSK, QPSK, 16QAM, and 64QAM.
OFDM modulation is then used to convert the signal from the frequency domain to the time-domain.
In the OFDM modulation, we use 64 FFT points and 16 points for cyclic prefix (CP). 
Channel coding was not implemented for data transmission, as it can be independently assessed. 


Fig.~\ref{fig:phasenoisedis}(a) shows the constellation of the decoded signal at the receiver when QPSK modulation is used at the transmitter. 
The resulting EVM of the decoded signals is about $-7$ dB.
Apparently, the current mmWave testbed cannot support QPSK modulation transmission as the resulting EVM cannot meet the required threshold ($-10$ dB).  
We found that the root cause of the poor performance is the phase noise induced from mmWave frequency up-/down-conversion.


\noindent \textbf{Phase Noise Measurement.}
To measure the phase noise, we transmit a single-frequency signal (40 MHz) on the transmitter side and analyze the received signal on the receiver side. 
Fig.~\ref{fig:phasenoisedis}(b) shows the distribution of phase noise measured at the receiver. 
As shown in the figure, the phase noise fits a Gaussian distribution, with mean $0$ and standard deviation $0.26$.
With such strong phase noise, it is easy to see why the signal decoded at the receiver significantly deviates from the original signal from the transmitter. 
Therefore, it is imperative to develop a scheme to address the phase noise problem. 

\section{Phase Noise Cancellation}

\begin{figure}
	\centering
	\includegraphics[width=2.25in]{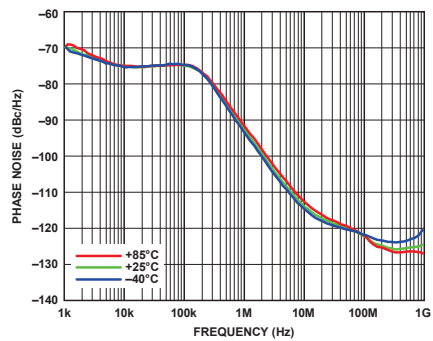}\vspace{-0.05in}
	\caption{Phase noise of reference clock from \cite{pncadi}.}\vspace{-0.15in}
	\label{fig:phase_noise_hmc}
\end{figure}

Phase noise is not limited to our testbed. 
It is an outstanding problem for generic mmWave communication systems due to the high oscillation frequency of mmWave transceiver. 
For example, Fig.~\ref{fig:phase_noise_hmc} shows the phase noise measured on another COTS mmWave transceiver ADI HMC6300/HMC6301 chipset \cite{pncadi}. 
It can be seen that the phase noise can be greater than -90~dBc/Hz at 1~MHz offset from the center frequency. 
Such significant phase noise poses a grand challenge in the design of mmWave communication systems.

The design of our PNC scheme is based on our observation that phase noise generated by frequency up-/down-conversion is not white.
As shown in Fig.~\ref{fig:phasenoisedis}(c), the dominant power density spectrum is located at low frequency ($< 1$ MHz).
Hence, if we can cancel the components of low-frequency phase noise, the performance of the mmWave testbed will be significantly improved. 
Guided by this idea, we aim to develop a scheme to estimate and mitigate the low-frequency phase noise at the receiver.
In what follows, we will first present a signal model in the presence of phase noise and then present our scheme in detail.
 
\subsection{Signal Model with Phase Noise}

\begin{figure}
	\centering
	\includegraphics[width=1\linewidth]{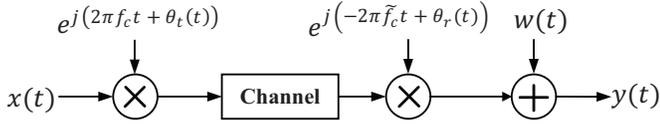}
	\caption{Signal flow model in the presence of phase noise.}\vspace{-0.15in}
	\label{fig:signalmodel}
\end{figure}

We consider the baseband signal flow as shown in Fig.~\ref{fig:signalmodel}.
On the transmitter side, the baseband signal $x(t)$ is up-converted to the carrier frequency $f_c$, which comes with phase noise $\theta_t(t)$. 
The RF signal is transmitted over-the-air channel, which is considered to be a multipath channel. 
On the receiver side, the received RF signal is down-converted to baseband signal using a local oscillator of frequency $\tilde{f}_c$, which comes with phase noise $\theta_r(t)$. 
As the frequency offset (the difference between $f_c$ and $\tilde{f}_c$) can be estimated and compensated at the receiver, we do not consider it in our model. 
Therefore, the baseband signal model can be written as:
\begin{equation}
y(n) = p_r(n)\sum_{l = 0}^{L - 1} h(l) p_t(n-l)x(n-l) + w(n),
~~
0 \le n < N,
\label{eq:signal_model1}
\end{equation}
where $y(n)$ is the received discrete signal at the receiver; $x(n)$ is the discrete transmit signal at the transmitter; $h(l)$ is the channel response with $L$ being the number of taps; $p_t(n) = e^{j \theta_t(n)}$ is the transmitter-side phase noise; $p_r(n) = e^{j \theta_r(n)}$ is the receiver-side phase noise, respectively; $w(n)$ is the signal noise. 
Since the bandwidth of phase noise is much smaller than the bandwidth of signal, we can consider 
$p_t(n-l) \approx p_t(n)$ for $0 \le l < L$.
Then, the baseband signal model can be rewritten as:
\begin{equation}
y(n) = p(n)\sum_{l = 0}^{L - 1} h(l) x(n-l) + w(n),
~~
0 \le n < N,
\label{eq:signal_model2}
\end{equation}
where $p(n) = p_r(n) p_t(n)$ is the combined phase noise, $\forall n$.

The baseband signal flow model in (\ref{eq:signal_model2}) indicates that, if we can estimate and mitigate the phase noise $p(n)$, the signal flow model in mmWave communications will be equivalent to that in conventional communication systems (e.g., 2.4 GHz carrier frequency with negligible phase noise).
As a result, the received signals at the receiver can be decoded using conventional solutions. 
Therefore, we develop a PNC scheme that can be applied to the mmWave testbed.

\subsection{A Phase Noise Cancellation Scheme}

We assume that the mmWave system uses OFDM modulation. 
Each OFDM symbol has $N$ subcarriers (FFT points).
A sequence of consecutive OFDM symbols constitutes a frame, which has several preamble OFDM symbols at the beginning for synchronization and channel estimation.
We assume that each OFDM symbol is perfectly synchronized in the time domain and frequency offset has been compensated. 
In such an OFDM-based mmWave system, we enable phase noise estimation and cancellation by inserting a reference signal on a subcarrier in each OFDM symbol.

\begin{figure}
	\centering
	\includegraphics[width=1\linewidth]{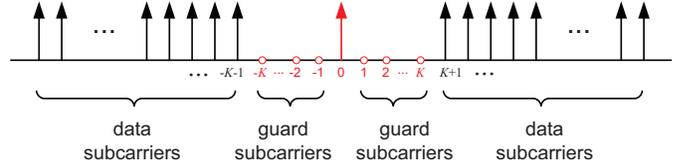}
	\caption{Illustration of pilot and guard subcarriers in an OFDM symbol.}\vspace{-0.15in}
	\label{fig:pilot_ofdm}
\end{figure}

\noindent
\textbf{Transmitter-Side Operations.}  
At the transmitter, instead of using all valid subcarriers for data transmission, we use one subcarrier for pilot and several subcarriers next to that subcarrier for guard bands, as illustrated in Fig.~\ref{fig:pilot_ofdm}.
The pilot signal is set to 1. 
The subcarrier selected for the pilot can be determined based on the system requirement. 
In our testbed, we use the DC subcarrier (subcarrier indexed 0) for the pilot, and use $K$ subcarriers for guard interval on each side (see Fig.~\ref{fig:pilot_ofdm}). 
Experimental results show that $K=3$ is sufficient to achieve decent performance. 
It is worth pointing out that in our testbed, the baseband signal is converted to IF signal using digital processing.
Hence, DC subcarrier can be safely used for phase pilot.
In mmWave systems without digital up-/down-conversion, DC subcarrier should not be used for phase pilot and another subcarrier (e.g., subcarrier 1 or -1) can be used instead.

\begin{figure}
	\centering
	\includegraphics[width=1\linewidth]{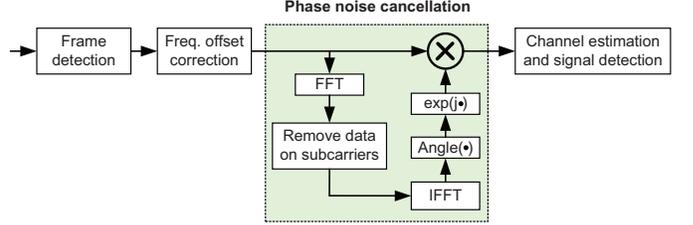}
	\caption{Phase noise cancellation algorithm at the receiver.}\vspace{-0.15in}
	\label{fig:phase_noise_cancellation}
\end{figure}

\begin{figure*}
	\begin{subfigure}[b]{2.25in}
		\centering
		\includegraphics[width=2.35in, height=1.6in]{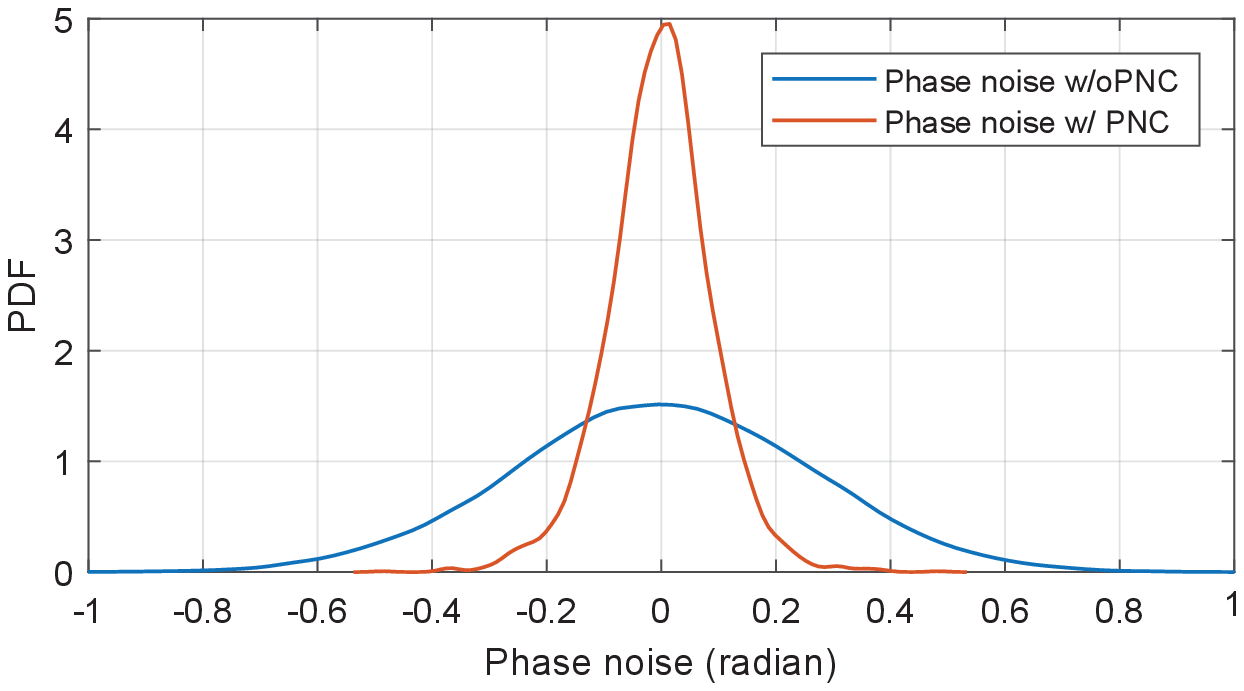}
		\caption{The distribution of phase noise measured on the testbed with and without PNC.}
		\label{fig:phasenoisecomp2}
	\end{subfigure}
	~~
	\begin{subfigure}[b]{2.5in}
		\centering
		\includegraphics[width=2.5in, height=1.6in]{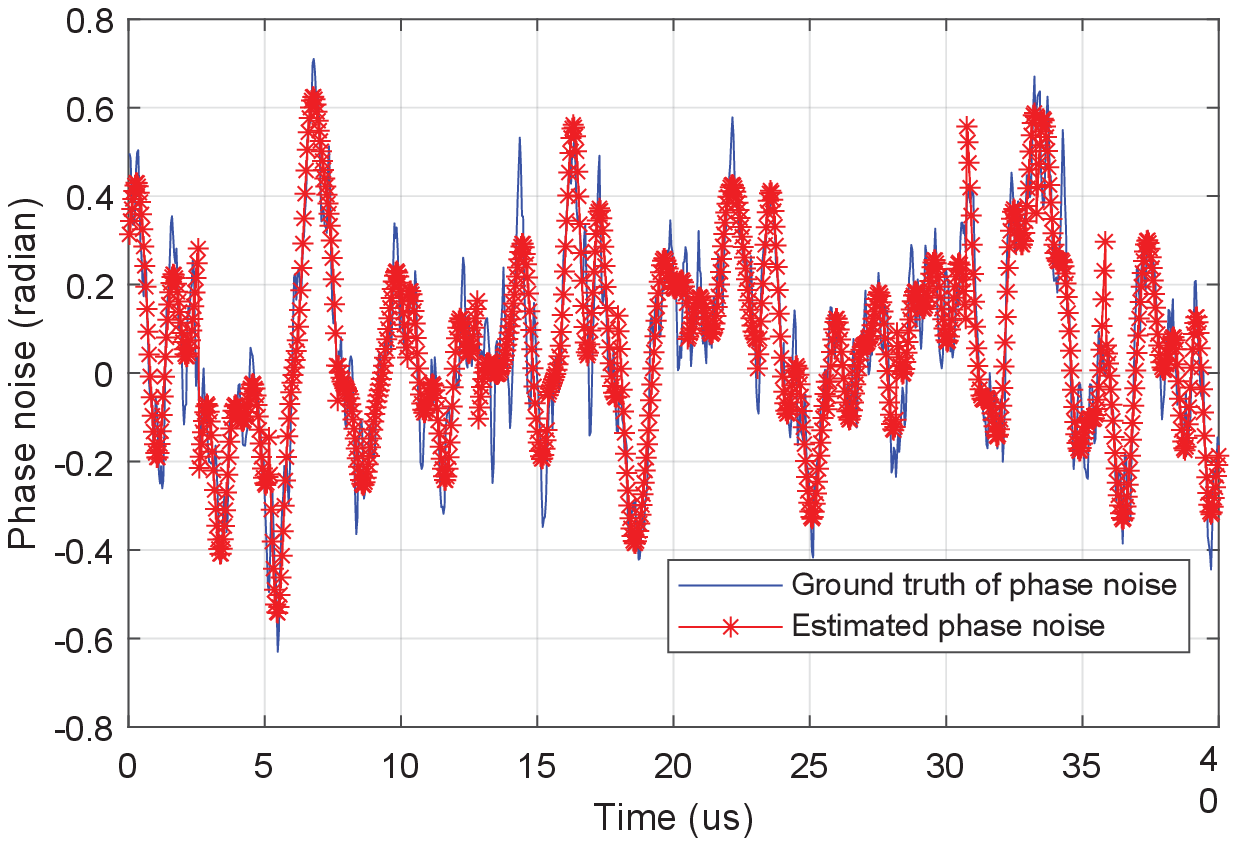}
		\caption{Ground truth of phase noise versus estimated phase noise at the receiver.}
		\label{fig:phasenoisecomp1}
	\end{subfigure}
	~~	
	\begin{subfigure}[b]{2.1in}
		\centering
		\includegraphics[width=2in, height=1.6in]{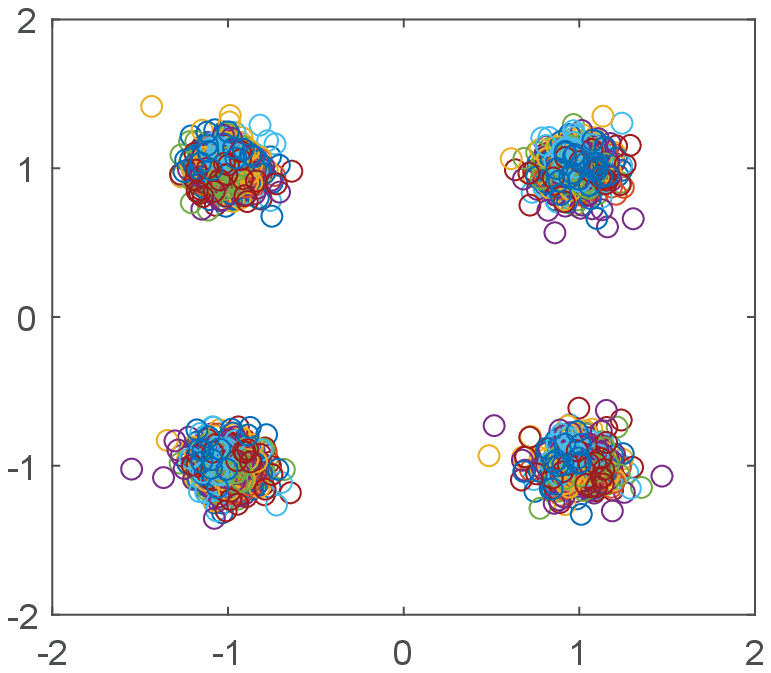}
		\caption{Constellation of decoded signal at the receiver with PNC (EVM = -19.5 dB).}
		\label{fig:constellationwpnc}
	\end{subfigure}	
	\vspace{-0.2in}	
	\caption{Experimental results from our mmWave testbed when the proposed PNC scheme is applied.}
	\vspace{-0.2in}
	\label{fig:mmwave_pnc}
\end{figure*}

\noindent
\textbf{Receiver-Side Operations.}  
At the receiver, we leverage the pilot signal and the guard subcarriers (null subcarriers) to estimate and cancel the phase noise for each individual OFDM symbol, as shown in Fig.~\ref{fig:phase_noise_cancellation}.
For the time-domain sample sequence $y(n)$ in (\ref{eq:signal_model2}), we perform FFT operation to obtain the data on the subcarriers, which we denote as $[Y(0), Y(1), \cdots, Y(N-1)]$.
Then, we nullify the subcarriers that are used to carry the payload and keep the data on DC and guard subcarriers (i.e., subcarriers indexed from $-K$ to $K$).
By denoting the resulting frequency-domain data as $[\tilde{Y}(0), \tilde{Y}(1), \cdots, \tilde{Y}(N-1)]$, we have
$\tilde{Y}(k) = Y(k)$ for $0 \le k \le K$ and $N-K \le k < N$
and
$\tilde{Y}(k) = 0$ otherwise.
After nullifying the payload subcarriers, we then take IFFT operation to convert the data from the frequency domain back to the time domain. 
Ideally, since the payload data has been nullified, the resulting time-domain data should purely come from the phase noise (if we neglect other imperfections such as thermal noise and circuit nonlinearity). 
Denote the resulting time-domain data as $[p(0), p(1), \cdots, p(N-1)]$.
Then, we cancel the phase noise for this OFDM symbol in the time domain by letting:
$r(n) = y(n) \cdot \exp(-j{\angle p(n))}$ for $0 \le n < N$, where $r(n)$ is the time-domain signal after phase noise cancellation and $N$ is the number of samples in an OFDM symbol.
After phase noise cancellation, the resulting time-domain signal $r(n)$ is fed to the blocks of channel estimation and signal detection for data recovery, as shown in Fig.~\ref{fig:phase_noise_cancellation}.

For the proposed PNC scheme, two remarks are in order.

\textit{Remark 1.}  
The proposed PNC scheme comes at a cost of extra bandwidth as it exclusively occupies a subset of subcarriers.
This cost can be justified by the large bandwidth available on mmWave frequencies.
As the bandwidth of signal (multiple GHz) is much larger than that of phase noise (typically 1 or several MHz), the number of guard subcarriers (i.e., $K$) is negligible compared the total number of subcarriers. 
As a guideline, the value of $K$ can be set to $B_{pn}/{\Delta f}$, where $B_{pn}$ is the bandwidth of phase noise and $\Delta f$ is the subcarrier spacing.

\textit{Remark 2.}  
Compared to other time-domain or frequency-domain PNC schemes \cite{dhananjay2015iris,zhang2015iterative}, our PNC scheme is amenable to practical implementation. 
We have implemented this PNC scheme on our testbed.
The increased computational complexity is marginal when running in real time.

\section{Experimental Results}

We have implemented the proposed PNC scheme on our mmWave testbed. 
In what follows, we first study the performance of the PNC scheme and then examine the performance of the mmWave testbed when PNC is used.

\noindent \textbf{Phase Noise.}
To quantify the effectiveness of our proposed PNC scheme, we measure the phase noise at the receiver with and without PNC. 
Fig.~\ref{fig:mmwave_pnc}(a) shows the distribution of phase noise measured on our mmWave testbed with and without using our proposed PNC scheme.
From the distribution, it is evident to see that our proposed PNC scheme can significantly reduce the phase noise.
Using the proposed PNC scheme, the standard deviation of the phase noise is reduced from 0.26 radian to 0.09 radian.
To see how the PNC scheme works, we present the ground truth of phase noise and the estimated phase noise in Fig.~\ref{fig:mmwave_pnc}(b).
We can see that the estimated phase noise in our PNC scheme is pretty accurate.
After subtracting the estimated phase noise, the residual phase noise is much smaller compared to the original phase noise.

\noindent \textbf{Constellation and EVM.}
We now evaluate the performance of our mmWave testbed when PNC is used. 
We conduct OFDM signal transmission using the parameters described in Section~\ref{sec:exp_obs} (i.e., 64 FFT points, 25 MHz bandwidth, and IEEE 802.11ac legacy frame structure). 
For the PNC scheme, we set the guard interval to 3 subcarriers (i.e., $K=3$).
We collect the decoded signal at the receiver and plot the constellation diagram in Fig.~\ref{fig:mmwave_pnc}(c).
Comparing the constellation diagrams in Fig.~\ref{fig:mmwave_pnc}(c) and Fig.~\ref{fig:phasenoisedis}(a), it can be seen that our proposed PNC scheme can significantly improve the performance of the mmWave testbed. 
Experimental measurement shows that the proposed PNC scheme improves the EVM of the decoded signal from -8 dB to -20 dB on average.

\noindent \textbf{Real-Time Video Streaming.}
As an application, we implemented video streaming on our testbed to demonstrate its applicability in practice.
As shown in Fig.~\ref{fig:practiaclsettingmmwave}, we feed video data packet to the transmitter, which then assembles the video data packet using a simplified version of IEEE 802.11 legacy PHY (no MCS and no channel coding) and sends the RF signal into the air on 58.32 GHz.
The mmWave receiver down-converts the received RF signal and decodes the video data packet using the signal processing modules shown in Fig.~\ref{fig:phase_noise_cancellation}.
For this application, we observed that the video can be played smoothly at the receiver, achieving an average EVM of -19.5 dB. 
A demo of video streaming on our mmWave testbed can be found in \cite{videodemo}.

\section{Discussions on Testbed Limitations}
While our mmWave testbed can support real-time video streaming and other applications, it suffers from three limitations, which we described as follows.

\noindent \textbf{Narrow Bandwidth.}
Our mmWave testbed uses USRP N210 and GNU Radio for signal processing and protocol implementation to ease its use.
As such, the bandwidth of our testbed is limited to 25 MHz. 
By using USRP X310, the bandwidth can be improved to 160 MHz, which is still insufficient for some mmWave applications.
Therefore, our testbed is positioned for narrow-band experiments and applications.

\cutoff{
\noindent \textbf{Phase Noise.}
Future mmWave communication systems (e.g., 802.11ay) will probably be required to support 256-QAM.
The proposed PNC scheme can support up to 16-QAM and is envisioned to be incapable of supporting 256-QAM (even with further optimization).
To fundamentally resolve the phase noise issue and to support high order of modulation, advanced semiconductor technology is needed for stable oscillation frequency generation.
}

\noindent \textbf{Horn Antenna.}
Since mmWave communication systems need to overcome the high path loss, the design and application of phased-array antenna are important research topics in this area. 
However, our testbed comes with a horn antenna which can only pour energy to one direction. 
This limits its applications in the research of beamforming and energy steering.

Despite these limitations, this easy-to-use mmWave testbed is expected to have many applications ranging from PHY validation to RF sensing on mmWave frequencies. 
 
\section{Related Work}

\noindent \textbf{Existing mmWave Testbeds.}
As mmWave is considered as a key technology for 5G, some mmWave testbeds have already been developed for researchers to conduct experiments on mmWave frequencies. 
In \cite{saha2019x60}, a programmable testbed called X60 was developed to support PHY and MAC reconfigurations for communications on mmWave frequencies. 
This testbed was built using National Instruments' millimeter-wave transceiver system and SiBeam's 12-element phased antenna arrays.
\cutoff{It can support over 2 GHz instantaneous channel bandwidth. 
On this testbed, the baseband signal is generated inside a NI PXIe-1085 PXI Express chassis, which can be programmed using LabVIEW.}
Despite its versatility, this mmWave testbed is extremely expensive (over \$200,000) and is not affordable for most research groups. 
In \cite{zhang2016openmili}, a 60 GHz software radio platform called OpenMili was developed to support broadband transmission on mmWave frequencies. 
It integrates hardware components including FPGA-based baseband processing unit, 60 GHz TX/RX RF frontends, and a custom-designed phased-array antenna. 
Particularly, the phase noise problem on this platform was addressed through hardware reclocking by using external high-precision oscillators to drive Tx/Rx RF frontends. 
Despite its large bandwidth and reconfigurability, reproducing this platform requires knowledge of mmWave circuit components and expertise in FPGA programming, which may pose a technical barrier for some researchers.

In addition to the above heavy-duty mmWave testbeds, lightweight and easy-to-use ones have also been explored to enable rapid prototyping for mmWave experimentation. 
In \cite{zetterberg2015open}, a mmWave testbed was built using USRP N210, TX RF frontend (Hittite HMC6000LP711E), and RX RF frontend (Hittite HMC6001LP711E). 
While this testbed is easy to use for general researchers, it relies on external clocks (Vectron VCC6-QCE-285M7140000) to address the phase noise problem. 
A similar methodology has been used to build a mmWave demo in \cite{arnold2015spectrum}, where USRP devices are used for baseband signal processing and SIVERSIMA converters to transfer the IF signal to the mm-wave RF band. 
However, it seems that the devices (60 GHz up/down converters) are not available on the market, making it hard to assemble the testbed for general researchers. 
Moreover, it is not clear how this mmWave testbed addressed the phase noise problem.


\noindent
\textbf{Phase Noise Cancellation.}
Phase noise is a critical problem in the design of high-performance mmWave communication systems. 
Coarsely speaking, this problem can be mitigated through two different approaches: developing high-quality oscillators and employing phase noise cancellation techniques.
As these two approaches advance in parallel, the focus of our work is on the development of phase noise cancellation techniques. 
In \cite{dhananjay2015iris}, the phase noise was studied on mmWave testbed and a new ramp model that fits the
experimental observations about phase noise process. 
In \cite{zhang2015iterative}, an iterative channel estimation scheme was proposed for phase noise compensation. 
However, this scheme was developed for single-carrier frequency-domain equalization (SC-FDE) mmWave systems and may not work for OFDM mmWave systems. 
Moreover, the proposed scheme is of high computational complexity. 
Our work differs from existing efforts by developing a PNC scheme that is amenable to practical implementation.

\section{Conclusions}
\vspace*{-0.025in}
This paper presents a reconfigurable mmWave testbed assembled using COTS devices and open-source software package. 
To address the notorious phase noise problem, we developed a PNC scheme that takes advantage of the pilot signal inserted at the transmitter to estimate and mitigate the low-frequency components of phase noise at the receiver.
We implemented a simplified version of IEEE 802.11 legacy PHY on our mmWave testbed.
Experimental results show that the testbed can support real-time video streaming using OFDM modulation, reaching an average of EVM -20 dB.
We hope that our testbed provides a reference for other research groups to build their own mmWave testbeds for the prototyping and validation of new research ideas.

\vspace*{-0.05in}
\section*{Acknowledgment}
\vspace*{-0.025in}
A.~Quadri's and H.~Zeng's research was supported in part by the NSF under grant CNS-1717840.
Y.T.~Hou's research was supported in part by NSF under grants CNS-1617634 and CNS-1642873.

\vspace*{-0.1in}
\bibliography{references}
\bibliographystyle{ieeetr}
\vspace*{-0.75in}
\end{document}